\documentclass[aps,pra,reprint,superscriptaddress,floatfix]{revtex4-2}
\usepackage{amsmath,amssymb,graphicx,bm}
\usepackage[hidelinks]{hyperref}
\newcommand{\ii}{\mathrm{i}}
\newcommand{\eps}{\epsilon}
\begin{document}
	\raggedbottom
	\title{Recurrent excitations of vector Akhmediev breathers from initial modulations\\with unequal frequencies}
	\author{Muchen Dong}
	\author{Lei Wang}
	\email{50901924@ncepu.edu.cn}
	\affiliation{School of Mathematics and Physics, North China Electric Power University, Beijing 102206, People's Republic of China}
	\author{Chong Liu}
	\email{chongliu@nwu.edu.cn}
	\affiliation{School of Physics, Northwest University, Xi'an 710127, China}
	\affiliation{Shaanxi Key Laboratory for Theoretical Physics Frontiers, Xi'an 710127, China}
	\affiliation{Peng Huanwu Center for Fundamental Theory, Xi'an 710127, China}
	\affiliation{Fundamental Discipline Research Center for Quantum Science and Technology of Shaanxi Province, Xi'an 710127, China}
	\begin{abstract}
		We study the modulation instability (MI) phenomena in the Manakov system induced by initial modulations with different frequencies. Evolution from two distinct pairs of initial sidebands exhibits successive growth-decay cycles in both the degenerate and nondegenerate regimes. Corresponding exact multi-Akhmediev breathers are constructed to describe the wave evolution and Fourier spectra. With a common modulation frequency, new sidebands appear only at its higher harmonics. By contrast, four-wave mixing between initial modulations at different frequencies generates sidebands at the sum, difference and at higher order combinations of the two input frequencies, where unstable sideband pairs can give rise to distinct Akhmediev breathers. An unexpected feature of the evolution is that breathers with lower MI gains can be excited before those with higher gains. To explain this phenomenon, we use matched asymptotic expansions to show that the order of recurrent excitations depends jointly on the MI gains and the evolution of growing mode coefficients over successive cycles.
	\end{abstract}
	\maketitle
	\section{Introduction}
	
	Modulation instability is a fundamental nonlinear phenomenon in optics~\cite{Tai1986}, hydrodynamics~\cite{BenjaminFeir1967}, and Bose--Einstein condensates~\cite{Nguyen2017}. Weak periodic modulations of a plane wave grow exponentially when their frequencies lie within an instability band. In the nonlinear Schr\"odinger equation (NLSE), the Akhmediev breather (AB) describes the full growth-decay cycle of such modulations~\cite{Akhmediev1986}. During this cycle, the energy transfer from the central mode to the sidebands and back provides an analytic description of Fermi--Pasta--Ulam--Tsingou recurrence, which has been observed in optical experiments~\cite{VanSimaeys2001,Mussot2018,Pierangeli2018}. Four-wave mixing generates higher harmonics during the evolution. If several harmonics lie within the instability band, they can be excited and produce more complex growth-decay dynamics described by exact multi-AB solutions~\cite{Erkintalo2011}.
	
	Compared with the scalar case, nonlinear coupling between two wave components gives rise to richer MI dynamics. As a vector extension of the scalar NLSE, the Manakov system~\cite{Manakov1974} describes such coupled waves in optical fibers~\cite{Agrawal2013}, Bose--Einstein condensates~\cite{Kevrekidis2008}, and crossing seas~\cite{Onorato2006}. Vector ABs can have different wave profiles in the two components and exhibit asymmetry between positive and negative modes in their physical spectra~\cite{Chen2021Asymmetry}. In the nondegenerate regime, two distinct ABs can coexist at the same characteristic frequency with equal or unequal MI gains. A single pair of initial sidebands can simultaneously excite both nondegenerate ABs~\cite{Liu2022Nondegenerate}. Moreover, higher harmonics generated under equal-frequency modulation can grow and lead to frequency jumps across stable gaps~\cite{Chen2023HigherOrder}. Numerical simulations further show that the modulation of only one component can excite rogue waves and superregular breathers~\cite{Chen2026SingleComponent}.
	
	Nonlinear evolution becomes more complicated once the two components are initially modulated at different frequencies. The difficulty lies in the fact that four-wave mixing generates sidebands at the sum and difference frequencies and at higher-order frequency combinations, in addition to the higher harmonics reported for equal-frequency modulation~\cite{Chen2023HigherOrder}. Moreover, the existing approaches (both the Darboux transform~\cite{Matveev1991} and Hirota's bilinear method~\cite{Hirota2004Direct}) can only be used to construct the exact AB solutions that admit the same characteristic frequency for two components~\cite{Chen2021Asymmetry,Liu2022Nondegenerate,LingZhao2019Homoclinic}. Consequently, how to describe exactly the nonlinear recurrent evolution from initial modulations with unequal frequencies remains completely unexplored. On the other hand, independent modulation of the two components is common and is a standard experimental preparation. Optical experiments have demonstrated nonlinear interactions between two waves initially modulated at different frequencies~\cite{Emami2024Doppler,Okawachi2012Crosstalk}. Vector AB experiments have also observed recurrence and spectral asymmetry with an input arrangement that allows the modulation frequencies of the two components to differ~\cite{Liu2025Experimental}. Therefore, a gap remains between the experiments and the corresponding descriptions of the evolution dynamics, whether analytical or numerical. In particular, the key issues are to identify which generated sidebands develop into growth-decay cycles and to determine the order of excitations.
	
	To fill this gap, in our work, we study recurrent AB excitations from two modulations with different frequencies in the focusing Manakov system. In both the degenerate and nondegenerate regimes, the sidebands generated at mixed frequencies develop into new growth-decay cycles. Corresponding exact multi-AB solutions are constructed to describe the numerical wave evolution and physical spectra. An unusual phenomenon emerges in which breathers with lower MI gains can be excited before those with higher gains. We use matched asymptotic expansions~\cite{Grinevich2018Matched} to explain the observed excitation sequence. For initial modulations at different frequencies, the resulting asymptotic criteria combine the MI growth rates with the evolution of growing mode coefficients to predict the order of excitation events.
	
	\section{Vector ABs, MI and physical spectra}
	The focusing Manakov equations~\cite{Manakov1974} can be written in the following dimensionless form
	\begin{equation}
		\ii q_{j,t}+q_{j,xx}+2\bigl(|q_1|^2+|q_2|^2\bigr)q_j=0, \qquad j=1,2 .
		\label{eq:manakov}
	\end{equation}
	Here $q_1(x,t)$ and $q_2(x,t)$ are the two nonlinearly coupled components of the vector wave field. The physical interpretation of $x$ and $t$ depends on the application. In fiber optics, $t$ represents the normalized propagation distance, while $x$ is the normalized retarded time in a frame moving at the group velocity~\cite{Agrawal2013}. In Bose--Einstein condensates, $x$ and $t$ represent position and time, respectively~\cite{Kevrekidis2008}.
	
	We consider the plane-wave background
	\begin{equation}
		q_j^{(0)}(x,t)=a_j e^{\ii\theta_j},\qquad
		\theta_j=b_jx+ \bigl[2(a_1^2+a_2^2)-b_j^2\bigr]t,
	\end{equation}
	with $j=1,2$, where $a_j$ and $b_j\in\mathbb R$ are the background amplitudes and wavenumbers, respectively. A Darboux transformation yields a first-order vector AB of the form~\cite{Chen2021Asymmetry,Liu2022Nondegenerate}
	\begin{equation}
		\frac{q_j^{\mathrm{AB}}}{q_j^{(0)}}= \frac{
			(h_j^*/h_j)e^\Gamma+e^{-\Gamma} +\varrho\bigl(h_j^*e^{\ii\Theta} +h_j^{-1}e^{-\ii\Theta}\bigr)
		}{
			2(\cosh\Gamma+\varrho\cos\Theta)
		}.
		\label{eq:AB-field-model}
	\end{equation}
	Here $*$ denotes complex conjugation, and
	\begin{align}
		\Gamma&=2\kappa\operatorname{Im}\chi\,(t-t_0), \nonumber\\
		\Theta&=\kappa\bigl[x-x_0+ (2\operatorname{Re}\chi+\kappa)(t-t_0)\bigr].
		\label{eq:AB-arguments-model}
	\end{align}
	The characteristic frequency $\kappa$ determines the spatial period $2\pi/\kappa$. Here $x-x_0$ and $t-t_0$ are the shifted spatial and temporal variables, respectively, with $x_0,t_0\in\mathbb{R}$ specifying the position of the breather. The other factors are
	\[
	\varrho=\frac{2\operatorname{Im}\chi} {\sqrt{4(\operatorname{Im}\chi)^2+\kappa^2}}, \qquad
	h_j=\frac{\chi+b_j}{\chi+\kappa+b_j},
	\]
	where $\chi$ satisfies
	\begin{equation}
		1+\sum_{j=1}^{2} \frac{a_j^2}{(\chi+b_j)(\chi+\kappa+b_j)}=0.
		\label{eq:AB-root-model}
	\end{equation}
	Without loss of generality, a Galilean transformation allows us to set $b_1=-b_2=\beta$. For simplicity, we take $a_1=a_2=a>0$ in what follows.
	
\begin{figure}[t]
		\includegraphics[width=\columnwidth]{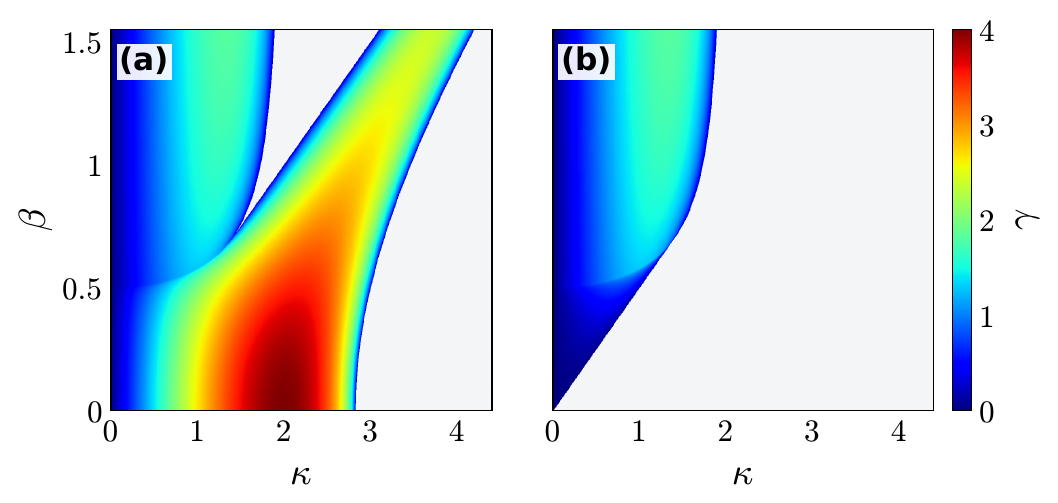}
		\caption{MI growth rate maps for $a=1$. (a) and (b) show the larger and smaller growth rates $\gamma_1$ and $\gamma_2$, respectively. Both maps span $0\leq\kappa\leq4.4$ and $0\leq\beta\leq1.55$. Light gray denotes the region of zero growth rate.}
		\label{fig:maps}
	\end{figure}

	The MI growth rate follows directly from the exact AB solution. Equation~\eqref{eq:AB-root-model} generically has four roots, with nonreal roots occurring in complex-conjugate pairs. At fixed $\kappa>0$, each distinct root $\chi$ in the upper half-plane can label one growing AB branch. Equation~\eqref{eq:AB-arguments-model} then gives
	\begin{equation}
		\gamma=2\kappa\operatorname{Im}\chi,\qquad
		\nu=\kappa\bigl(2\operatorname{Re}\chi+\kappa\bigr).
		\label{eq:AB-MI-rates-model}
	\end{equation}
	Expanding the AB solution about the plane-wave background as $t\to-\infty$ gives perturbation terms proportional to $e^{\gamma t\pm\ii(\kappa x+\nu t)}$. Thus $\gamma$ is the MI growth rate and $\kappa$ corresponds to the modulation frequency. The MI growth rate maps in Fig.~\ref{fig:maps} are obtained by solving Eq.~\eqref{eq:AB-root-model} and evaluating the growth rates in Eq.~\eqref{eq:AB-MI-rates-model}. A positive gain in Fig.~\ref{fig:maps}(a) and zero gain in Fig.~\ref{fig:maps}(b) identify the degenerate regime, with one growing AB branch. Where both gains are positive, two ABs coexist at the same modulation frequency $\kappa$, giving the nondegenerate regime. These ABs share the common spatial period $2\pi/\kappa$ and may have equal or unequal MI gains. At equal gains, the modulation patterns propagate with equal and opposite phase velocities $-\nu/\kappa$. When two growing branches coexist at the same $\kappa$, we denote their roots by $\chi_\eta(\kappa)$, with $\eta\in\{+,-\}$. For unequal gains, $+$ and $-$ label the branches with the larger and smaller MI growth rates, respectively. For equal gains, the labels follow the sign of $\nu$.
	
	To study the excitation of vector breathers, we solve the system~\eqref{eq:manakov} numerically using the split-step Fourier method. The simulations start from two plane-wave background components, each with a weak modulation at a different frequency,
	\begin{equation}
		q_{j}(x,0)=a[1+\eps\cos(\omega_j x)]e^{\ii b_jx},
		\label{eq:input}
	\end{equation}
	where $\omega_1\ne\omega_2$ and $0<\eps\ll1$. The common $\eps$ sets the same amplitude for the initial modulations. For convenience, we choose commensurate modulation frequencies and write their ratio as $\omega_1/\omega_2=n_1/n_2$, where $n_1$ and $n_2$ are coprime positive integers. Then the fundamental modulation frequency can be defined as $\omega_0=\omega_1/n_1=\omega_2/n_2$, so that $\omega_j=n_j\omega_0$ for $j=1,2$. Set $\kappa_\ell=\ell\omega_0$, $\ell=1,2,\ldots$. If $\kappa_\ell$ supports a single MI branch, we denote its root by $\chi_\ell$. When two branches coexist, we write $\chi_{\ell,\eta}=\chi_\eta(\kappa_\ell)$ for their roots. For a fixed root $\chi$ at modulation frequency $\kappa$, let $q_j(\chi,\kappa)$ denote the $j$-th component of the fundamental AB in Eq.~\eqref{eq:AB-field-model}.
	
	Removing the plane-wave phase gives the envelopes $U_j=q_j e^{-\ii\theta_j}$, which share the spatial period $L=2\pi/\omega_0$. The physical spectrum of each component expands during breather growth and contracts during decay, reflecting the transfer of energy from the central mode to the sidebands and back~\cite{Chen2021Asymmetry}. We calculate the Fourier coefficients and their normalized amplitudes as
	\begin{align}
		\widehat U_{j,n}(t) &=\frac{1}{L}\int_0^L U_j(x,t)e^{-\ii n\omega_0x}\,\mathrm{d} x, \\
		A_{j,n}(t) &=\frac{|\widehat U_{j,n}(t)|}{a}, \qquad n\in\mathbb Z.
	\end{align}
	In each component, the Fourier frequencies $n\omega_0$ are measured relative to the carrier, with $n=0$ denoting the central mode. The initial sidebands in component $j$ are at $\pm\omega_j=\pm n_j\omega_0$ ($n=\pm n_j$), each with Fourier amplitude $a\eps/2$. For equal background amplitudes and opposite carrier wavenumbers, exchanging the wave components and reversing $x$ leaves the background unchanged and interchanges the initial modulation frequencies. It follows that
	\begin{equation}
		A_{1,n}^{(\omega_1,\omega_2)}(t) =A_{2,-n}^{(\omega_2,\omega_1)}(t).
	\end{equation}
	Thus, interchanging the two modulation frequencies maps mode $n$ in the spectrum of the second component to mode $-n$ in the spectrum of the first component. Therefore, we plot the first-component spectrum at selected times to compare the sideband amplitudes below.
	
	\section{Breather excitation in the degenerate regime}
	\label{sec:degenerate}
	
\begin{figure}[t]
		\includegraphics[width=\columnwidth]{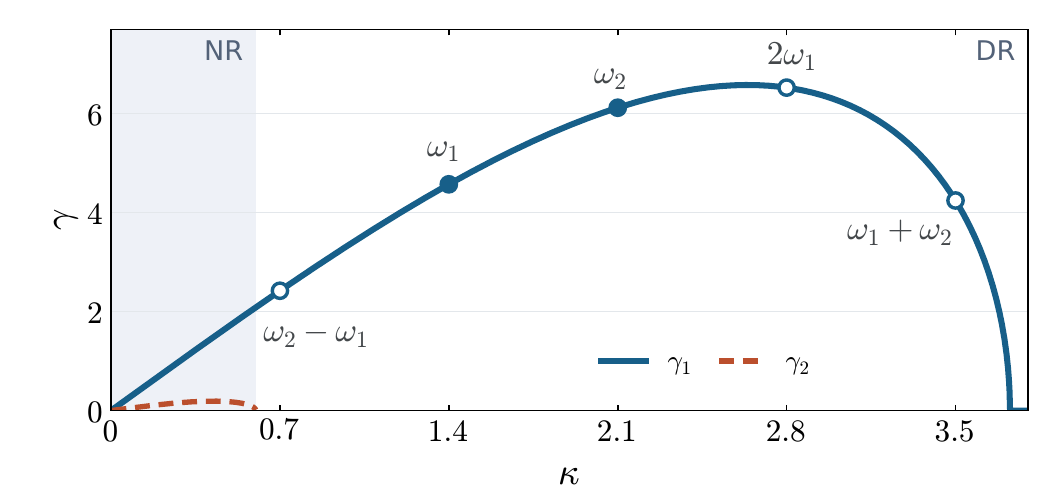}
		\caption{MI growth rate spectrum for $a=1.3$ and $\beta=0.3$ with $(\omega_1,\omega_2)=(1.4,2.1)$. Solid circles mark the initial modulation frequencies and open circles mark other unstable frequencies. Within the instability band, the shaded interval $0<\kappa<0.6$ is the nondegenerate region (NR) and the unshaded part is the degenerate region (DR).}
		\label{fig:cutS}
	\end{figure}

	We first examine the growth-decay cycles that develop in the degenerate regime, where each unstable frequency supports only one growing AB branch. Figure~\ref{fig:cutS} shows the MI growth rate spectrum for this case. Both initial modulation frequencies lie within the instability band, together with $2\omega_1$, $\omega_1+\omega_2$, and $\omega_2-\omega_1$ generated by four-wave mixing. These five frequencies are the first five integer multiples of $\omega_0$ and all lie in the degenerate regime. A second growing branch exists on the interval \(0<\kappa<0.6\), below the first allowed frequency \(\omega_0=0.7\). The second harmonic $2\omega_2=6\omega_0$ and other higher-order combinations of frequencies lie outside the instability band. The largest gain among these frequencies occurs at $2\omega_1$, while the smallest occurs at $\omega_2-\omega_1$, respectively. Notably, neither sideband pair is present in the initial spectrum.
	
	Four-wave mixing is the fundamental mechanism to generate the additional sidebands~\cite{McKinstrie2004FourWaveMixing,Chen2023HigherOrder}. Initially, each component contributes a modulation at $\omega_j$ to the total intensity $I=|q_1|^2+|q_2|^2$. This intensity contains both imposed frequencies and enters the evolution of each component through the nonlinear term $2Iq_j$ in Eq.~\eqref{eq:manakov}. These products generate sidebands at the sum and difference frequencies through cross terms between the two modulations, while the square of either modulation generates its second harmonic. Specifically, these frequency combinations follow from
	\begin{equation}
		\begin{aligned}
			2\cos(\omega_1x)\cos(\omega_2x) &=\cos[(\omega_1+\omega_2)x]+\cos[(\omega_2-\omega_1)x],\\
			\cos^2(\omega_jx) &=\frac12[1+\cos(2\omega_jx)].
		\end{aligned}
	\end{equation}
	Consequently, the initial sidebands have amplitudes of order $\eps$, whereas those at $\omega_1+\omega_2$, $\omega_2-\omega_1$, and $2\omega_1$ first appear at order $\eps^2$. Coupling through the total intensity allows either wave component to develop sidebands at a frequency initially imposed on the other.
	
\begin{figure*}[p]
		\includegraphics[width=0.93\textwidth]{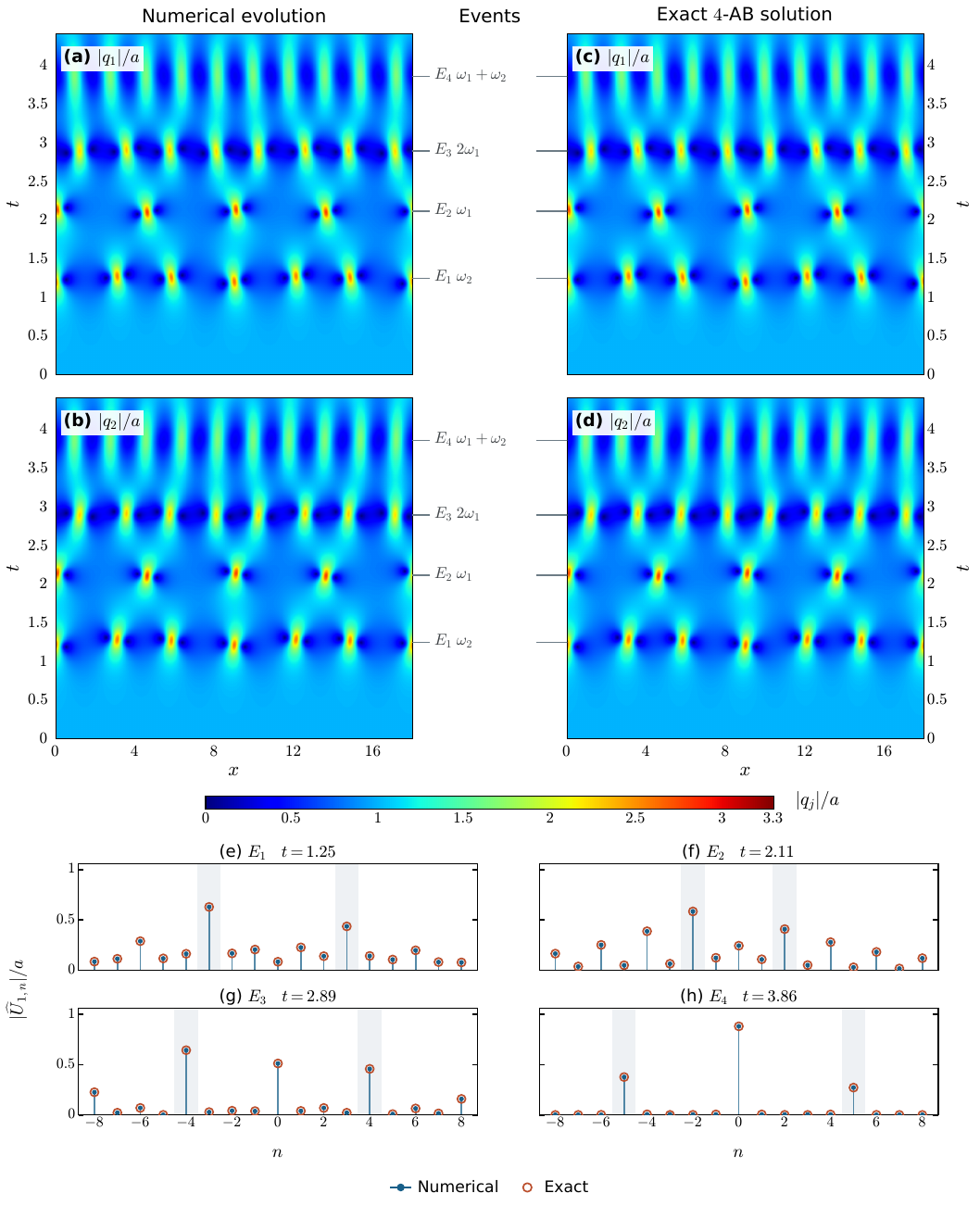}
		\caption{The numerical evolution and exact multi-AB solution in the degenerate regime. The simulation starts from the initial condition in Eq.~\eqref{eq:input} with $a=1.3$, $\beta=0.3$, $\eps=2\times10^{-3}$, and $(\omega_1,\omega_2)=(1.4,2.1)$, giving $\omega_0=0.7$. Figs.~\ref{fig:S}(a,b) show the numerical evolutions, and (c,d) show the exact 4-AB solution. The $(x_0,t_0)$ pairs of the constituent ABs are $(0.03,\,1.78)$, $(-2.10,\,2.04)$, $(0.21,\,2.70)$, and $(-0.79,\,3.46)$ in the order $E_1$--$E_4$. Figs.~\ref{fig:S}(e)--(h) compare the physical spectra $A_{1,n}$ of the first component. Filled points and stems show the numerical spectra, and orange open circles show the exact 4-AB spectra. Each spectrum is evaluated at the predicted excitation time of the labeled event. For an event at $\kappa_\ell=\ell\omega_0$, the gray bands mark the Fourier modes $n=\pm\ell$.}
		\label{fig:S}
	\end{figure*}

	Figure~\ref{fig:S} shows the MI dynamics for the initial condition in Eq.~\eqref{eq:input} with the parameter values given in the caption. Figs.~\ref{fig:S}(a,b) show the numerical evolution of the two wave components, while Figs.~\ref{fig:S}(c,d) present the exact 4-AB solution for comparison. Figs.~\ref{fig:S}(e)--(h) compare the numerical and analytical Fourier spectra of the first component at selected times. The first two cycles, $E_1$ and $E_2$, are associated with two ABs $q_j(\chi_3,\omega_2)$ and $q_j(\chi_2,\omega_1)$, respectively. The sidebands at $\pm\omega_2$ dominate the spectrum in Fig.~\ref{fig:S}(e). The lower frequency $\omega_1$ accounts for the longer spatial period at $E_2$, where the sidebands at $\pm\omega_1$ become prominent in Fig.~\ref{fig:S}(f). Since both initial modulations enter at order $\eps$, the larger MI gain at $\omega_2$ favors excitation of the corresponding AB before the AB at $\omega_1$.
	
	Higher harmonics have already been visible during the first two cycles. The spectrum at $E_2$ contains nonzero sidebands at $\pm2\omega_1$. These sidebands are also visible at $E_1$. A distinct growth--decay cycle develops at $E_3$ with half the spatial period of the preceding cycle at $E_2$. The strongest sideband pair in the first-component spectrum lies at $\pm 2\omega_1$ [Fig.~\ref{fig:S}(g)]. These features identify the AB $q_j(\chi_4,2\omega_1)$ as the constituent of $E_3$. For the chosen frequencies, the second harmonic $2\omega_1$ coincides with the combination frequency $2\omega_2-\omega_1$. The second harmonic process $\omega_1+\omega_1$ contributes to these sidebands at order $\eps^2$, whereas mixing via $\omega_2+\omega_2-\omega_1$ can only contribute at order $\eps^3$. Therefore, the second harmonic dominates this growth-decay cycle rather than third-order mixing. Although this frequency has the largest MI growth rate among the unstable frequencies, its excitation is not the first to appear. Thus, the MI growth rates alone do not determine the excitation order, which will be examined quantitatively in Sec.~IV.
	
	At $E_4$, four-wave mixing excites the AB $q_j(\chi_5,\omega_1+\omega_2)$ at a frequency outside the harmonic series of either input. Unequal modulation frequencies thus open a new route to breather excitation at a distinct sum frequency, whereas under equal-frequency modulation the same mixing process feeds only the second harmonic. Among the observed cycles, $E_4$ has the highest modulation frequency and hence the shortest spatial period $2\pi/(\omega_1+\omega_2)$, giving rise to the most closely spaced patterns in $x$. The corresponding sideband also dominates the spectrum in Fig.~\ref{fig:S}(h). It is notable that the sidebands outside the harmonic series of $\omega_1+\omega_2$ carry little power. This spectrum thus more closely resembles that of an isolated AB at the same frequency. The difference frequency $\omega_2-\omega_1$ also has positive MI gain but does not undergo its first excitation within this time interval. It starts from a second-order contribution and has the smallest gain among the unstable frequencies according to Fig.~\ref{fig:cutS}. Therefore, its amplification takes longer.
	
	A distinctive feature of the Fourier spectra in Fig.~\ref{fig:S}(e)--(h) is that sidebands outside the harmonic series of the corresponding AB have smaller amplitudes in the later cycles. For an isolated AB with modulation frequency $\kappa$, the sidebands occur only at $\pm m\kappa$, $m=1,2,\ldots$. On either side of the central mode, the first order sideband has the largest amplitude, and higher order sidebands decrease geometrically~\cite{Chen2021Asymmetry}. At $E_1$, the spectrum also contains sidebands at frequencies that are not integer multiples of $\omega_2$, so it differs from that of a single AB with modulation frequency $\omega_2$. By $E_4$, these additional sidebands have become weak, and the spectrum approaches that of an isolated AB. Simulations over a longer propagation time show that these additional sidebands grow again during later cycles. Their decrease from $E_1$ to $E_4$ therefore represents a local trend within the observed evolution cycles. The amplitude of the central Fourier mode also increases from one compression to the next over these four cycles.
	
	Fig.~\ref{fig:S} shows that the exact 4-AB solution reproduces the evolution in both components, and its spectra agree closely with those obtained from the numerical simulation. Compared to the degenerate case, richer MI dynamics become possible in the nondegenerate regime as considered next, where the same modulation frequency can support two distinct AB branches.
	
	\section{Breather excitation in the nondegenerate regime and its order}
	\label{sec:order}
	
	We now consider the nondegenerate regime, in which two distinct MI branches can coexist at the same modulation frequency. A single pair of sidebands at that frequency can excite both corresponding ABs. To examine these richer dynamics, we choose parameters such that one of the initial modulation frequencies lies in the nondegenerate regime. Some frequencies generated by four-wave mixing also correspond to two different ABs even with different MI growth rates.
	
	\subsection{Successive excitation of AB branches}
	
	\begin{figure}[t]
		\includegraphics[width=\columnwidth]{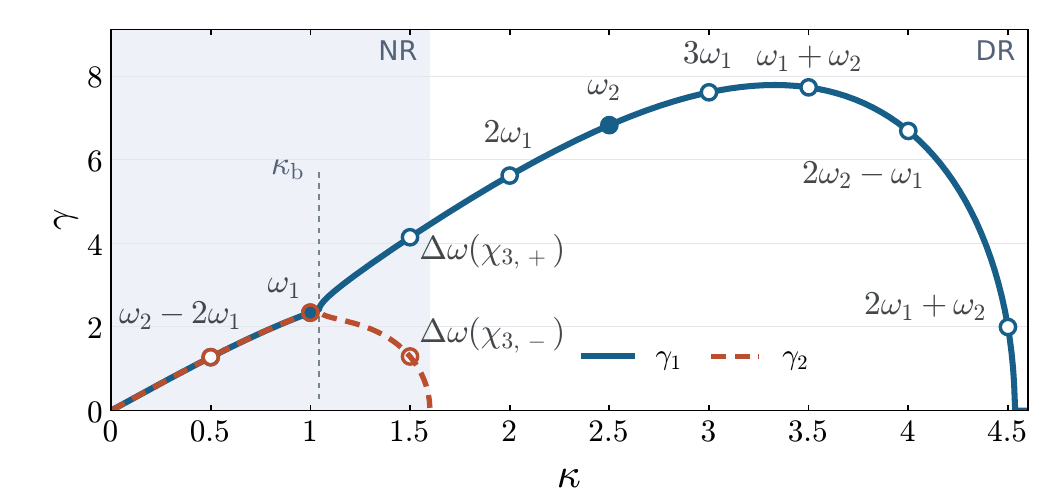}
		\caption{MI growth rate spectrum for $a=1.5$ and $\beta=0.8$ with $(\omega_1,\omega_2)=(1.0,2.5)$. Filled and open blue circles mark the imposed and other unstable grid frequencies, respectively. Orange circles mark the gain of a second growing branch where one exists. The dashed line marks $\kappa_{\mathrm b}$, where the two roots coalesce. At $\omega_1$, the roots $\chi_{2,-}$ and $\chi_{2,+}$ both give an MI gain of approximately $2.34$. At $\Delta\omega=\omega_2-\omega_1$, $\chi_{3,+}$ and $\chi_{3,-}$ give MI growth rates of approximately $4.15$ and $1.29$, respectively. Within the instability band, the shaded interval $0<\kappa<1.6$ is the nondegenerate region (NR) and the unshaded part is the degenerate region (DR).}
		\label{fig:cutV}
	\end{figure}
	
	Figure~\ref{fig:cutV} shows the MI growth rate spectrum for the nondegenerate case. The two branches in the shaded interval coalesce at $\kappa_{\mathrm b}$. The modulation frequencies $n\omega_0$, $n=1,\ldots,9$, lie in the instability band, whereas $2\omega_2=10\omega_0$ and all higher harmonics of $\omega_0$ lie outside. At $\omega_1$, the roots $\chi_{2,-}$ and $\chi_{2,+}$ yield the same MI growth rate, with $\nu<0$ and $\nu>0$, respectively. The corresponding nondegenerate ABs have phase velocities that are equal in magnitude and opposite in sign. For the same shifts $(x_0,t_0)$, reflection about $x=x_0$ maps $|q_j(\chi_{2,+},\omega_1)|$ onto $|q_{3-j}(\chi_{2,-},\omega_1)|$ for $j=1,2$. At $\kappa_3=\Delta\omega=\omega_2-\omega_1$, the roots $\chi_{3,+}$ and $\chi_{3,-}$ yield the larger and smaller MI growth rates, respectively. The two corresponding nondegenerate AB solutions generally have different amplitude profiles. However, here the same reflection acts within each branch $\eta=\pm$, relating $|q_1(\chi_{3,\eta},\Delta\omega)|$ to $|q_2(\chi_{3,\eta},\Delta\omega)|$. Only the difference-frequency branch with the larger gain is excited within the displayed interval.
	
	\begin{figure*}[p]
		\includegraphics[width=0.91\textwidth]{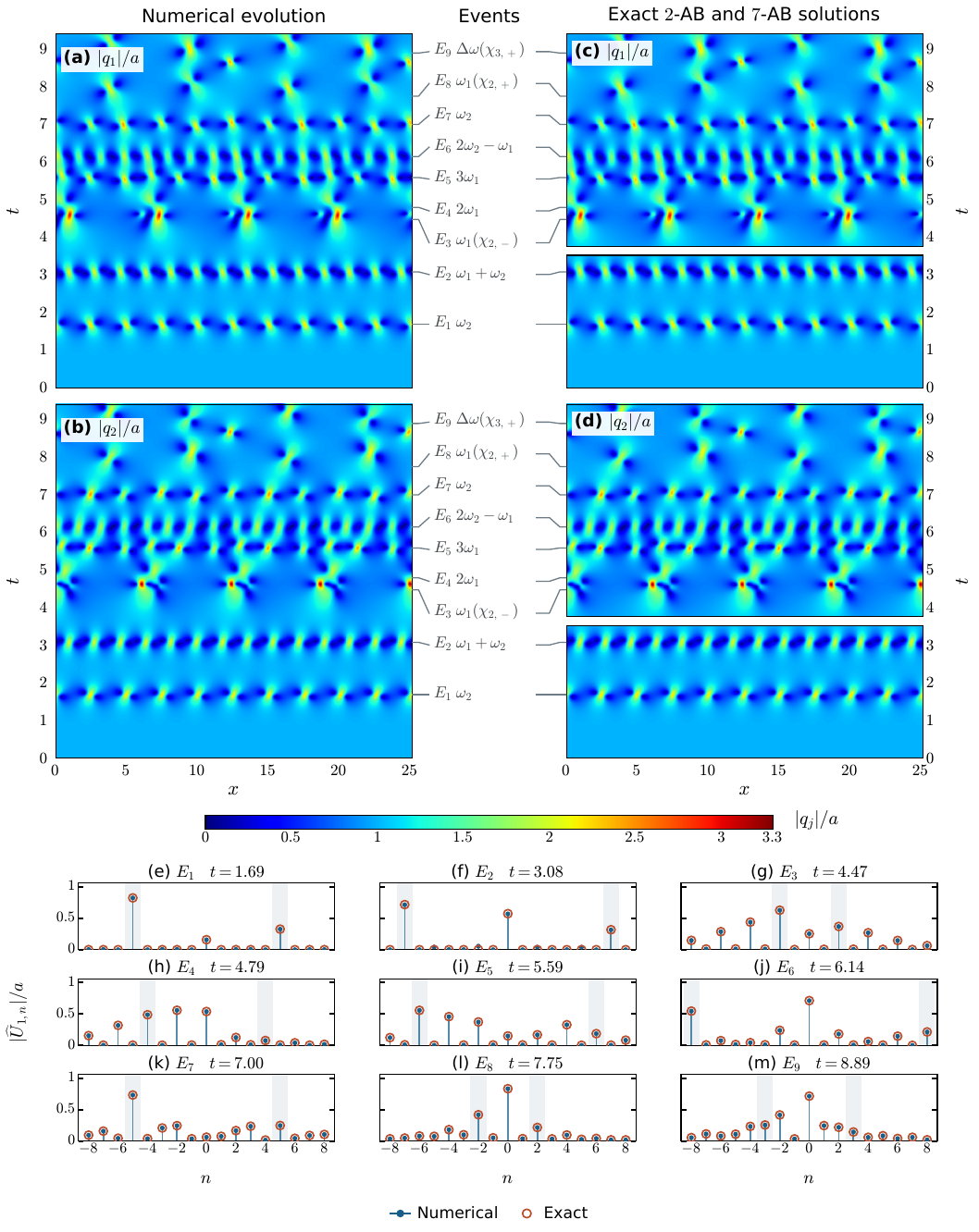}
		\caption{The numerical and exact multi-AB evolution in the nondegenerate regime. The simulation starts from the initial condition in Eq.~\eqref{eq:input} with $a=1.5$, $\beta=0.8$, $\eps=3.5\times10^{-5}$, and $(\omega_1,\omega_2)=(1.0,2.5)$, giving $\omega_0=0.5$. Figs. (a,b) show the numerical evolution and Figs. (c,d) show the corresponding exact 2-AB solution for cycles $E_1$ and $E_2$ and the exact 7-AB solution for cycles $E_3$--$E_9$. In event order, the $(x_0,t_0)$ pairs are $(-1.20,\,1.87)$ and $(-0.70,\,2.92)$ for the 2-AB solution, and $(-0.29,\,6.53)$, $(1.06,\,5.78)$, $(-0.11,\,5.93)$, $(0.49,\,6.08)$, $(0.37,\,6.49)$, $(-2.31,\,6.77)$, and $(0.86,\,7.27)$ for the 7-AB solution. Figs. (e)--(m) compare the numerical and analytical first-component spectra $A_{1,n}$ at the predicted excitation time of each event. For an event at $\kappa_\ell=\ell\omega_0$, the gray bands mark the modes $n=\pm\ell$.}
		\label{fig:V}
	\end{figure*}
	
	Figure~\ref{fig:V} shows the MI dynamics for the initial condition in Eq.~\eqref{eq:input} with the parameters given in the caption. Figs.~\ref{fig:V}(a,b) show the numerical evolution of the two wave components, and Figs.~\ref{fig:V}(c,d) present the corresponding exact multi-AB solutions. Figs.~\ref{fig:V}(e)--(m) compare the numerical and analytical Fourier spectra of the first wave component at selected times. The initial modulation at $\omega_2$ produces the first cycle, $E_1$, associated with $q_j(\chi_5,\omega_2)$. The sum-frequency sidebands are absent from the initial spectrum, yet their associated AB $q_j(\chi_7,\omega_1+\omega_2)$ is excited at $E_2$ even before the branch at the other initial frequency $\omega_1$. The exact 2-AB solution reproduces the numerical evolution of $E_1$ and $E_2$ in both components, with good agreement between the numerical and analytical physical spectra in Figs.~\ref{fig:V}(e,f).
	
	After that, the overlapping cycles $E_3$ and $E_4$ associated with $q_j(\chi_{2,-},\omega_1)$ and $q_j(\chi_4,2\omega_1)$ are excited, respectively. In both spectra in Figs.~\ref{fig:V}(g,h), the power of the sideband pair at $n=\pm2$ is largest among $n>0$. The overlap allows this pair associated with $E_3$ to dominate in $E_4$ as well, where its modulation frequency $2\omega_1$ actually corresponds to $n=\pm4$. Cycles $E_5$ and $E_6$ correspond to two ABs $q_j(\chi_6,3\omega_1)$ and $q_j(\chi_8,2\omega_2-\omega_1)$, respectively. Their spectra are shown in Figs.~\ref{fig:V}(i,j). Both branches are seeded at order $\eps^3$, and the earlier excitation at $3\omega_1$ is consistent with its larger MI gain in Fig.~\ref{fig:cutV}. The next excitation at $2\omega_2-\omega_1$ reveals another new mechanism of unequal-frequency excitation. Different orderings of the three frequency shifts $+\omega_2$, $+\omega_2$, and $-\omega_1$ give rise to distinct mixing paths. Each contributes a third-order modulation amplitude at this frequency. Two possible paths are
	\begin{equation}
		\begin{aligned}
			\omega_2 &\xrightarrow{+\omega_2}2\omega_2 \xrightarrow{-\omega_1}2\omega_2-\omega_1,\\
			\omega_2 &\xrightarrow{-\omega_1}\omega_2-\omega_1 \xrightarrow{+\omega_2}2\omega_2-\omega_1.
		\end{aligned}
	\end{equation}
	Notably, the first path passes through $2\omega_2$, which lies outside the instability band. Therefore, this intermediate stable frequency can still contribute to the modulation at $2\omega_2-\omega_1$. Different contributions to the same modulation will be added, and the first nonzero term determines the order and strength of the resulting modulation, as detailed in Appendix~\ref{app:seeds}. For unequal initial modulation amplitudes \(\eps_1\) and \(\eps_2\) in the two components, the third-order contributions at \(3\omega_1\) and \(2\omega_2-\omega_1\) are proportional to \(\eps_1^3\) and \(\eps_1\eps_2^2\), respectively. Varying the amplitude may induce more complex dynamics. We omit the details for brevity, since our method extends directly to this case.
	
	The branch with root $\chi_5$ at $\omega_2$, which is first excited at $E_1$, undergoes its second cycle at $E_7$. Its spectrum at $E_7$ is shown in Fig.~\ref{fig:V}(k). The pattern is slightly shifted in $x$ relative to $E_1$ owing to phase shifts from the preceding cycles. At $E_8$, the AB $q_j(\chi_{2,+},\omega_1)$ is excited for the first time with its spectrum shown in Fig.~\ref{fig:V}(l). Its partner $q_j(\chi_{2,-},\omega_1)$ has the same MI growth rate and reached its first compression at $E_3$. Thus the two nondegenerate ABs undergo their first compressions in separate cycles despite sharing the same modulation frequency and MI gain. This separation is a distinctive feature of the unequal frequency excitation considered here, whereas modulations with equal frequencies can only excite both ABs simultaneously in a nonlinear superposition~\cite{Liu2022Nondegenerate}. Their opposite phase velocities account for the opposite drifts at $E_3$ and $E_8$. At $E_9$, the difference-frequency AB $q_j(\chi_{3,+},\Delta\omega)$ with the larger MI growth rate reaches compression. Figure~\ref{fig:V}(m) shows the corresponding first-component spectrum. Between the cycles $E_8$ and $E_9$, the two ABs interact and produce rogue wave like peaks that repeat along $x$ in both wave components.
	
	The recurrence at $E_7$ involves the same Darboux spectral parameter as $E_1$, so the Darboux transformation used here cannot directly yield a single 9-AB representation. We therefore use exact 2-AB and 7-AB solutions for the early and later time intervals, respectively. Both the numerical evolution in the two components and the corresponding physical spectra agree well with the analytical results. The branches with roots $\chi_{1,+}$ and $\chi_{1,-}$ at $\omega_2-2\omega_1$ do not reach their first compression within the displayed interval. The same holds for the branch with root $\chi_9$ at $2\omega_1+\omega_2$ and the branch with root $\chi_{3,-}$ at $\Delta\omega$ corresponding to the smaller MI growth rate.
	
	\subsection{Asymptotic criteria for excitation order}
	
	In Fig.~\ref{fig:V}, the cycle $E_1$ at the initial frequency $\omega_2$ precedes the cycle $E_2$ at the sum frequency $\omega_1+\omega_2$, although the latter one corresponds to the larger MI growth rate as shown in Fig.~\ref{fig:cutV}. This observation suggests that the excitation order depends on more than the MI growth rates. Although a larger MI growth rate increases the amplification rate, higher order combinations of frequencies contribute at higher powers of $\eps$, so the corresponding modulations may take longer to develop into ABs. It is therefore important to clarify how these two factors act together. Besides, once the first excitation of $q_j(\chi_5,\omega_2)$ at $E_1$ is complete, the competition mechanism between its recurrence and the growth of sidebands at other frequencies remains unclear. To address these issues, we use matched asymptotic expansions~\cite{Grinevich2018Matched} to derive criteria for predicting the order of AB excitations. A similar approach has been used to generate asymptotic ABs from initial modulation with equal frequencies~\cite{Ling2026Asymptotic}.
	
	We now consider which AB branch is excited first. Both the MI growth rate of each branch and the initial amplitude of its growing mode are required. We use the Lax pair to determine this amplitude by matching the initial cosine modulations to the growing tail of an exact AB solution. Let $i$ label a fixed AB branch with modulation frequency $\kappa_i=\ell_i\omega_0$ and the selected root $\chi_i$ with $\operatorname{Im}\chi_i>0$ from Eq.~\eqref{eq:AB-root-model}. We denote its MI growth rate by $\gamma_i$. Then the matching in  Appendix~\ref{app:seeds} gives an initial coefficient of the growing mode in the form $\alpha_i^0=\eps^{p_i}G_i$, where $G_i$ is the complex factor combining all contributions from different mixing paths. Here $p_i$ equals the minimum number of terms from $\{\pm\omega_1,\pm\omega_2\}$ whose sum is $\kappa_{i}$. For the first cycle on branch $i$, we define the predicted excitation time as $T_i^{(0)}$. For an isolated AB, this is the center time $t_0$. When cycles overlap, the wave profile may peak at a different time. The predicted time difference between branches $i$ and $k$ is then
	\begin{equation}
		\Delta T_{ik}^{(0)} = T_i^{(0)}-T_k^{(0)}
		= \left(\frac{p_i}{\gamma_i}-\frac{p_k}{\gamma_k}\right) \log\left(\frac{1}{\eps}\right)+O(1).
		\label{eq:initial_clock}
	\end{equation}
	For sufficiently small $\eps$, $p_i/\gamma_i<p_k/\gamma_k$ gives $\Delta T_{ik}^{(0)}<0$, so the AB on branch $i$ is predicted to develop before the AB on branch $k$. When the two ratios are close or $\eps$ is not small enough, the subleading terms given in Appendix~\ref{app:reference} must be retained.
	
	Then we consider the order of subsequent excitation cycles. Let $N_m$ count the cycles already completed for the AB branch labeled $m$, and write $\bm N=(N_1,N_2,\dots)$. By Eq.~\eqref{eq:amplitude_update}, each completed cycle on branch $i$ multiplies its own growing mode coefficient by a factor of order $\eps^{2p_i}$, while cycles on other branches contribute factors of order unity. Starting from order $\eps^{p_i}$, therefore, the growing coefficient for the next excitation on branch $i$ is of order $\eps^{(2N_i+1)p_i}$. Let $T_i^{(\bm N)}$ denote the predicted excitation time of the next compression for branch $i$ after the cycles counted by $\bm N$. Then the time difference between branches $i$ and $k$ has the form
	\begin{equation}
		\Delta T_{ik}^{(\bm N)} = \left[ \frac{(2N_i+1)p_i}{\gamma_i} -\frac{(2N_k+1)p_k}{\gamma_k} \right]\log\left(\frac1\eps\right)+O(1).
		\label{eq:clock}
	\end{equation}
	If the coefficient of $\log(1/\eps)$ vanishes, high order corrections are required. Eq.~\eqref{eq:clock} reduces to Eq.~\eqref{eq:initial_clock} when $\bm N=\bm0$.
	
	For example, after the cycle $E_1$, neither $q_j(\chi_7,\omega_1+\omega_2)$ nor $q_j(\chi_{2,-},\omega_1)$ has been excited. Their growing mode coefficients start at second and first order in $\eps$, respectively. The AB $q_j(\chi_5,\omega_2)$ at $E_1$ has completed one cycle, so its recurrence enters the comparison with $p_i=1$ and $N_i=1$. Writing $\gamma_{\omega_1}$ for the common gain of the two nondegenerate branches at $\omega_1$, we find from Fig.~\ref{fig:cutV} that the coefficients of $\log(1/\eps)$ in Eq.~\eqref{eq:clock} satisfy
	\begin{equation}
		\frac{2}{\gamma_{\omega_1+\omega_2}} < \frac{1}{\gamma_{\omega_1}} < \frac{3}{\gamma_{\omega_2}}.
	\end{equation}
	For sufficiently small $\eps$, the first inequality shows that the larger gain at the sum frequency compensates for the higher order of $\eps$. The second inequality places a first compression at $\omega_1$ before the recurrence at $\omega_2$. This agrees with the relative order of $E_2$, $E_3$, and $E_7$ in Fig.~\ref{fig:V}. Before $E_3$, the predicted compression times for the two branches at $\omega_1$ have the same leading logarithmic term. Let $p$ and $q$ label the branches with roots $\chi_{2,-}$ and $\chi_{2,+}$, respectively. Both growing mode coefficients are of order $\eps$, and $\gamma_p=\gamma_q$, so the leading logarithmic terms coincide. For each branch $m$ involved in $E_1$ or $E_2$, the multipliers in Eq.~\eqref{eq:amplitude_update} satisfy $|\sigma_{pm}|=|\sigma_{qm}|$. The shifts due to these two cycles therefore cancel in the time difference, and Eq.~\eqref{eq:clock_full} gives
	\begin{equation}
		-\frac{\log|G_p|}{\gamma_p}+\frac{\log|G_q|}{\gamma_q} =\frac{1}{\gamma_p}\log\frac{|G_q|}{|G_p|}.
	\end{equation}
	With $G_p\simeq-5.575+3.697\ii$ and $G_q\simeq4.568-4.453\ii$, we have $|G_p|>|G_q|$. The matching thus predicts that $q_j(\chi_{2,-},\omega_1)$ reaches compression first, as observed at $E_3$. Since $|\sigma_{qp}|<1$, $E_3$ delays the predicted first excitation on branch $q$, and subsequent shifts place it at $E_8$.
	
	The criteria in Eqs.~\eqref{eq:initial_clock} and \eqref{eq:clock} assume weak initial modulation and small spectral gaps. For fixed cycle counts $\bm N$ and sufficiently small $\eps$, the branch with the smaller coefficient of $\log(1/\eps)$ reaches compression first. If the logarithmic coefficients coincide or the correction to the time difference is non-negligible compared with the logarithmic term, we must use the full expressions in Appendix~\ref{app:reference} to compare the predicted excitation times. When overlapping cycles leave no interval in which both tails are small, their order can be difficult to resolve in the wave evolution. We therefore consider the excitation process as a whole. The predicted excitation times of the individual cycles still guide the construction of the corresponding exact multi-AB solution.
	
	\section{Conclusions}
	
	In conclusion, we have studied recurrent AB excitations from initial modulations with unequal frequencies in the focusing Manakov system. Four-wave mixing generates sidebands at harmonics and at combinations of the two input frequencies. Unequal-frequency excitation in the nondegenerate regime illustrates three new mechanisms compared to equal-frequency excitation. First, the same sideband at high order combination of frequencies receives contributions from distinct mixing pathways, one of which passes through a linearly stable band. Second, the initially absent sum-frequency sidebands give rise to an AB cycle before the first excitation at one of the initial frequencies. Third, two nondegenerate ABs with the same modulation frequency and MI growth rate are excited in separate AB cycles. The numerical evolution and Fourier spectra in both regimes agree well with exact multi-AB solutions. Matched asymptotic expansions give the criteria for predicting the excitation sequence.
	
	\section*{Acknowledgments}
	The work is supported by the National Natural Science Foundation of China (NSFC) (Grants No.~12375002 and No.~12522501) and the Scientific Research Innovation Capability Support Project for Young Faculty (Grant No.~SRICSPYF-BS2025131).
	
	\appendix
	
	\section{Initial coefficients of the growing and decaying modes}
	\label{app:seeds}
	
	\subsection{Contributions of the initial modulations}
	
	Let $\bm q=(q_1,q_2)^T$, $J=\operatorname{diag}(1,0,0)$, and
	\[
	Q=\begin{pmatrix}
		0&\bm q^\dagger\\
		\bm q&0_{2\times2}
	\end{pmatrix},
	\]
	where $T$ and $\dagger$ denote transpose and conjugate transpose, respectively. The Lax pair for the focusing Manakov equations~\cite{Manakov1974} is given by
	\begin{align}
		\Psi_x&=\ii(\mu J+Q)\Psi,\qquad
		\Psi_t=B\Psi,\nonumber\\
		B&=\ii(\mu^2J+\mu Q)+[J,Q_x]-\ii[J,Q]Q.
	\end{align}
	Here $\Psi$ is the eigenfunction, $\mu$ is the complex spectral parameter and the commutator is defined by $[A,B]=AB-BA$. The gauge transformation $Y=\mathcal G^{-1}\Psi$, with $\mathcal G=\operatorname{diag}(1,e^{\ii\theta_1},e^{\ii\theta_2})$, gives $Y_x=\ii C Y$, where
	\begin{equation}
		C=\begin{pmatrix}
			\mu&\overline U_1&\overline U_2\\
			U_1&-b_1&0\\
			U_2&0&-b_2
		\end{pmatrix}.
	\end{equation}
	Here $U_j=q_j e^{-\ii\theta_j}$ is the wave field with its plane wave phase removed and the overbar denotes complex conjugation. Write $C_0(\mu)$ for this matrix at $U_1=U_2=a$. Its eigenvalues $z$ and corresponding eigenvectors $\bm r(z)$ satisfy
	\begin{equation}
		\mu(z)=z-\sum_{j=1}^2\frac{a^2}{z+b_j},\qquad
		\bm r(z)=\left(1,\frac{a}{z+b_1},\frac{a}{z+b_2}\right)^T.
	\end{equation}
	For $\eps=0$, the transformed time equation is $Y_t=\ii(C_0^2-4a^2I_3)Y$, where $I_3$ is the $3\times3$ identity matrix. The corresponding background eigenfunction is $Y(z)=\bm r(z)e^{\ii[zx+(z^2-4a^2)t]}$.
	
	For an AB branch labeled $i$ with modulation frequency $\kappa_i=\ell_i\omega_0$ and root $\chi_i$ of Eq.~\eqref{eq:AB-root-model}, set $v_i=\chi_i+\kappa_i$ and $\mu(\chi_i)=\mu(v_i)=\mu_i$. The roots can be written explicitly as
	\begin{equation}
		\chi_i=-\kappa_i/2+\Omega_i,
	\end{equation}
	with $\Omega_i^2=\beta^2+\kappa_i^2/4-a^2\pm\sqrt{D_i}$ and $D_i=a^4-4a^2\beta^2+\beta^2\kappa_i^2$. Here we select distinct roots with $\operatorname{Im}\Omega_i>0$ as the representatives. At $t=0$, the cosine modulations in Eq.~\eqref{eq:input} give $C-C_0=\eps\sum_{s\in\{\pm n_1,\pm n_2\}} V_s e^{\ii s\omega_0x}$, where
	\begin{equation}
		V_{\pm n_j}=\frac a2 \left(\bm e_1\bm e_{j+1}^T+\bm e_{j+1}\bm e_1^T\right), \quad j=1,2.
	\end{equation}
	Here $\bm e_1,\bm e_2,\bm e_3$ are the standard basis vectors in $\mathbb R^3$. Substitution of $Y=e^{\ii k_Bx}\sum_m\bm y_m e^{\ii m\omega_0x}$ ($m\in\mathbb Z$) into $Y_x=\ii C Y$, with Bloch wavenumber $k_B$ and Fourier coefficients $\bm y_m$, gives
	\begin{equation}
		(C_0-m\omega_0I_3)\bm y_m +\eps\sum_s V_s\bm y_{m-s}=k_B\bm y_m.
	\end{equation}
	Then we expand each $\bm y_m$ in the eigenbasis of $C_0(\mu_i)$ for $(m,z)=(0,\chi_i)$ and $(\ell_i,v_i)$ to obtain a homogeneous $2\times2$ linear system for the coefficient of $\bm r(\chi_i)$ in $\bm y_0$ and that of $\bm r(v_i)$ in $\bm y_{\ell_i}$. The background diagonal term is
	\begin{equation}
		\bm r(z)^T[C_0(\mu)-(k_B+m\omega_0)I_3]\bm r(z) =(\mu-\mu_i)-\mathcal N(z)(k_B-\chi_i),
	\end{equation}
	where $\mathcal N(z)=\mu'(z)=\bm r(z)^T\bm r(z)$. Eliminating all other components, including the remaining directions within these two blocks, gives the reduced $2\times2$ system. We use local coordinates $(\delta\mu,\delta k_B)$ centered at the point near $(\mu_i,\chi_i)$ where both diagonal entries of the reduced matrix vanish. Let $p_i$ be the minimum length of an ordered sequence of shifts $s_r\in\{\pm n_1,\pm n_2\}$ whose sum is $\ell_i$. The leading system can be given by
	\begin{equation}\label{eq:leading}
		\begin{pmatrix}
			\delta\mu-\mathcal N(\chi_i)\delta k_B & \eps^{p_i}S_i\\
			\eps^{p_i}S_i & \delta\mu-\mathcal N(v_i)\delta k_B
		\end{pmatrix}
		\begin{pmatrix}
			\frac{\bm r(\chi_i)^T\bm y_0}{\mathcal N(\chi_i)}\\[3pt]
			\frac{\bm r(v_i)^T\bm y_{\ell_i}}{\mathcal N(v_i)}
		\end{pmatrix} =\bm 0.
	\end{equation}
	For each such ordered sequence $(s_1,\ldots,s_{p_i})$, set $m_r=\sum_{h=1}^r s_h$ and $\bm w_0=\bm r(\chi_i)$. The coefficient $S_i$ in the off-diagonal term is given by
	\begin{align}
		\mathcal R_i(m)&=\left[(\chi_i+m\omega_0)I_3-C_0(\mu_i)\right]^{-1}, \nonumber\\
		\bm w_r&=\mathcal R_i(m_r)V_{s_r}\bm w_{r-1}, \qquad 1\le r<p_i,\nonumber\\
		S_i&=\sum_{s_1+\cdots+s_{p_i}=\ell_i} \bm r(v_i)^T V_{s_{p_i}}\bm w_{p_i-1}. 
	\end{align}
	The two off-diagonal terms have the same coefficient $S_i$ at order $\eps^{p_i}$, since $\mathcal R_i(m)^T=\mathcal R_i(m)$ and $V_{-s}=V_s^T$. If this coefficient vanishes, higher order corrections are required. In particular, here we assume that $\chi_i$ and $v_i$ are isolated, the intermediate blocks are invertible, and $\mathcal N(\chi_i)$ and $\mathcal N(v_i)$ are nonzero and distinct.
	
	\subsection{Initial coefficients and the spectral gap}
	
	For a single AB, let $e$ and $o$ denote $\chi_i$ and $v_i$ in either order. Up to a scalar factor, its Darboux vector is $\bm y=\bm r(e)+s\bm r(o)$, where $s$ is the relative complex weight and $|s|\ll1$. Define
	\[
	g_j(e)=\frac{\bar e+b_j}{e+b_j}, \qquad
	c(e,o)=\frac{e-\bar e}{o-\bar e}.
	\]
	Then the normalized Darboux field has the expansion
	\begin{equation}
		\frac{U_j}{a g_j(e)}-1 =-\frac{(o-e)c(e,o)}{o+b_j}s +\frac{(\bar o-\bar e)\overline{c(e,o)s}} {\bar e+b_j}+O(|s|^2).
	\end{equation}
	Let $\alpha_i$ and $\delta_i$ denote the complex coefficients of the growing and decaying AB tails on branch $i$, respectively. For the growing tail, take $(e,o)=(v_i,\chi_i)$ and $s=\alpha_i e^{\gamma_i t-\ii(\kappa_i x+\nu_i t)}$, with $c_i^+=c(v_i,\chi_i)$. For the decaying tail, take $(e,o)=(\chi_i,v_i)$ and $s=\delta_i e^{-\gamma_i t+\ii(\kappa_i x+\nu_i t)}$, with $c_i^-=c(\chi_i,v_i)$. Here $\gamma_i$ and $\nu_i$ are given by Eq.~\eqref{eq:AB-MI-rates-model}. Set $\Delta\mathcal N_i=\mathcal N(\chi_i)-\mathcal N(v_i)$. Then we take $U_j(x,0)/g_j(e)-a$ as the initial perturbation and derive a system analogous to Eq.~\eqref{eq:leading}. The upper-right and lower-left terms are $\alpha_i\kappa_i c_i^+\Delta\mathcal N_i$ and $\delta_i\kappa_i c_i^-\Delta\mathcal N_i$, respectively. Matching each term to $\eps^{p_i}S_i$ determines the leading coefficients of initial modes $\alpha_i^0=\eps^{p_i}G_i$ and $\delta_i^0=\eps^{p_i}H_i$, with
	\begin{equation}
		G_i=\frac{S_i}{\kappa_i c_i^+\Delta \mathcal N_i}, \qquad
		H_i=\frac{S_i}{\kappa_i c_i^-\Delta \mathcal N_i}.
		\label{eq:seed_normalization}
	\end{equation}
	
	We next identify a quantity that is conserved across successive cycles. Since $Y$ is nontrivial, the determinant of Eq.~\eqref{eq:leading} must vanish, giving
	\begin{equation}
		[\delta\mu-\mathcal N(\chi_i)\delta k_B] [\delta\mu-\mathcal N(v_i)\delta k_B]-\eps^{2p_i}S_i^2=0. 
	\end{equation}
	At the nearby Floquet branch points $\mu_i^\pm$, the two roots in $\delta k_B$ coincide, so the discriminant vanishes,
	\begin{equation}
		(\Delta\mathcal N_i)^2(\delta\mu)^2 +4\mathcal N(\chi_i)\mathcal N(v_i)\eps^{2p_i}S_i^2=0.
	\end{equation}
	Using Eq.~\eqref{eq:seed_normalization}, we then obtain the spectral gap
	\begin{equation}
		(\mu_i^+-\mu_i^-)^2 =-16\mathcal N(\chi_i)\mathcal N(v_i)\kappa_i^2c_i^+c_i^-P_i +o(\eps^{2p_i}),
	\end{equation}
	where $P_i=\alpha_i^0\delta_i^0=\eps^{2p_i}G_iH_i$. Since the gauged Lax time matrix $\widetilde B=\mathcal G^{-1}B\mathcal G-\mathcal G^{-1}\mathcal G_t$ is $L$-periodic in $x$, the monodromy matrix $\mathcal M$ satisfies $\mathcal M_t=[\widetilde B(0,t;\mu),\mathcal M]$. Hence the branch points $\mu_i^\pm$ are independent of $t$. Therefore, the coefficients of the growing and decaying modes in each matching interval satisfy $\alpha_i\delta_i=P_i+o(\eps^{2p_i})$.
	
	\section{Excitation times and exact multi-AB solutions}
	\label{app:reference}
	
	\subsection{Predicted excitation times of AB cycles}
	
	For an AB cycle on another branch $m$ with $\mu_m\ne\mu_i$, let $D_w(\mu)$ denote the one-fold Darboux matrix
	\begin{equation}
		D_w(\mu)=I_3+ \frac{\bar\mu_m-\mu_m}{\mu-\bar\mu_m} \frac{\bm r(w)\bm r(w)^\dagger} {\bm r(w)^\dagger\bm r(w)},
	\end{equation}
	with $w\in\{\chi_m,v_m\}$ and $\mu_m=\mu(\chi_m)=\mu(v_m)$. Its action on $\bm r(z)$ at $\mu=\mu_i$ is
	\begin{equation}
		D_w(\mu_i)\bm r(z) =\frac{z-w}{z-\bar w} \operatorname{diag}(1,g_1(w),g_2(w))\bm r(z),
	\end{equation}
	where $z\in\{\chi_i,v_i\}$. Thus, the weight of $\bm r(\chi_i)$ relative to that of $\bm r(v_i)$ is multiplied by
	\begin{equation}
		\mathcal E_i(w)=\frac{\chi_i-w}{v_i-w} \frac{v_i-\bar w}{\chi_i-\bar w}, \qquad
		\sigma_{im}=\frac{\mathcal E_i(\chi_m)}{\mathcal E_i(v_m)}, \quad i\ne m.
		\label{eq:relative_weight}
	\end{equation}
	Therefore, across an AB cycle on branch $m$, the coefficients of the growing and decaying modes on each branch $i\ne m$ obey the evolution law $(\alpha_i,\delta_i)\mapsto(\sigma_{im}\alpha_i,\delta_i/\sigma_{im})$. Note that factoring $s$ out of $\bm r(v_m)+s\bm r(\chi_m)$ gives $\bm r(v_m)$ the relative weight $s^{-1}$ after a cycle. Therefore, if the growing mode coefficient before compression is $\alpha_m^{\mathrm{in}}$, the decaying one afterward is $\delta_m^{\mathrm{out}}=1/\alpha_m^{\mathrm{in}}$. In the outgoing matching interval, $\alpha_m^{\mathrm{out}}\delta_m^{\mathrm{out}}
	=P_m+o(\eps^{2p_m})$, so $\alpha_m^{\mathrm{out}}=P_m\alpha_m^{\mathrm{in}}$ to the leading order. Let $N_m$ count the completed cycles on branch $m$ and set $\bm N=(N_1,N_2,\ldots)$. We have the general evolution law
	\begin{equation}
		\alpha_i(\bm N) =\alpha_i^0P_i^{N_i} \prod_{m\ne i}\sigma_{im}^{N_m}, \qquad
		\delta_i(\bm N)=\frac{P_i}{\alpha_i(\bm N)}.
		\label{eq:amplitude_update}
	\end{equation}
	
	Now we define the predicted excitation time $T_i^{(\bm N)}$ by $|\alpha_i(\bm N)|e^{\gamma_iT_i^{(\bm N)}}=1$. For two branches after the same completed cycles, $\Delta T_{ik}^{(\bm N)}=T_i^{(\bm N)}-T_k^{(\bm N)}$ is given by
	\begin{equation}
		\Delta T_{ik}^{(\bm N)} = -\frac{\log|\alpha_i(\bm N)|}{\gamma_i} +\frac{\log|\alpha_k(\bm N)|}{\gamma_k}. 
	\end{equation}
	At $\bm N=\bm0$, this becomes
	\begin{equation}
		\Delta T_{ik}^{(0)} = \left(\frac{p_i}{\gamma_i} -\frac{p_k}{\gamma_k}\right)\log(1/\eps)-\frac{\log|G_i|}{\gamma_i} +\frac{\log|G_k|}{\gamma_k}.
		\label{eq:initial_clock_full}
	\end{equation}
	Substituting Eq.~\eqref{eq:amplitude_update} into the expression for $\Delta T_{ik}^{(\bm N)}$ gives
	\begin{equation}
		\begin{split}
			\Delta T_{ik}^{(\bm N)} ={}&\Delta T_{ik}^{(0)} -\frac{ N_i\log|P_i| +\sum_{m\ne i}N_m\log|\sigma_{im}| }{\gamma_i} \\
			&+\frac{ N_k\log|P_k| +\sum_{m\ne k}N_m\log|\sigma_{km}| }{\gamma_k}.
		\end{split}
		\label{eq:clock_full}
	\end{equation}
	Since $\log|P_i|=-2p_i\log(1/\eps)+\log|G_iH_i|$, Eqs.~\eqref{eq:initial_clock_full} and \eqref{eq:clock_full} yield Eqs.~\eqref{eq:initial_clock} and \eqref{eq:clock}. The $O(1)$ terms must be retained when the logarithmic coefficients coincide. Even when these coefficients differ, the $O(1)$ correction to $\Delta T_{ik}^{(\bm N)}$ may still be significant if $\eps$ is not small enough. 
	
	\subsection{Construction of exact multi-AB solutions}
	
	We construct an exact $M$-AB solution to describe $M$ consecutive excitation cycles on branches with distinct spectral parameters. For $i=1,\ldots,M$, the background Darboux vector takes the form
	\begin{equation}
		\bm y_i^{[0]}=\bm r(v_i)+s_i\bm r(\chi_i), \qquad
		s_i=d_i e^{\gamma_i t-\ii(\kappa_i x+\nu_i t)},
	\end{equation}
	with a complex weight $d_i$. To determine each weight, choose one of the $M$ cycles as the reference and denote its branch by $h$. Let $N_{*,m}$ count the cycles on branch $m$ that precede the reference cycle in the predicted sequence, and write $\bm N_*=(N_{*,1},N_{*,2},\dots)$. We use $\alpha_i$ and $\delta_i$ for the coefficients in Eq.~\eqref{eq:amplitude_update} evaluated at $\bm N_*$.
	
	For each branch $i$, let $\{e_i,o_i\}=\{\chi_i,v_i\}$. When matching the tail of the selected AB, we normalize the Darboux vector so that the coefficient of $\bm r(e_i)$ is one. If the cycle precedes the reference cycle in the predicted sequence, we match its decaying tail and take $(e_i,o_i)=(\chi_i,v_i)$. For the reference cycle and the subsequent selected cycles, we match the growing tail and take $(e_i,o_i)=(v_i,\chi_i)$. In particular, $e_h=v_h$. The Darboux transformations on the other $M-1$ branches multiply the weight of $\bm r(o_i)$ by a factor $F_i$. From Eq.~\eqref{eq:relative_weight}, we obtain
	\begin{equation}
		F_i=\prod_{\substack{l=1\\l\ne i}}^M \frac{o_i-e_l}{e_i-e_l} \frac{e_i-\bar e_l}{o_i-\bar e_l}.
	\end{equation}
	Then the matching gives $d_i=\alpha_i/F_i$ when $e_i=v_i$ and $d_i=F_i/\delta_i$ when $e_i=\chi_i$.
	
	With the weights $d_i$ fixed, we start from $U_j^{[0]}=a$ and apply $M$ successive Darboux transformations~\cite{Matveev1991,Chen2023HigherOrder} as follows
	\begin{align}
		P^{[p]} &=\frac{\bm y_p^{[p-1]}(\bm y_p^{[p-1]})^\dagger} {(\bm y_p^{[p-1]})^\dagger\bm y_p^{[p-1]}},\qquad
		D_p(\mu) =I_3+\frac{\bar\mu_p-\mu_p}{\mu-\bar\mu_p}P^{[p]}, \nonumber\\
		\bm y_i^{[p]} &=D_p(\mu_i)\bm y_i^{[p-1]},\qquad i>p, \nonumber\\
		U_j^{[p]} &=U_j^{[p-1]}+(\bar\mu_p-\mu_p)(P^{[p]})_{j+1,1}.\label{eq:mdarboux}
	\end{align}
	The resulting $M$-AB field is independent of the choice of reference branch $h$. The phase factor accumulated over the preceding cycles is
	\begin{equation}
		\Phi_j= \prod_m\left[\frac{g_j(\chi_m)}{g_j(v_m)}\right]^{N_{*,m}}.
		\label{eq:phase}
	\end{equation}
	To evaluate the background, take $\bm y_i^{[0]}=\bm r(e_i)$ for all $i$. Equation~\eqref{eq:mdarboux} then gives $a\prod_{i=1}^{M}g_j(e_i)$ in component $j$. After restoring the accumulated phase factor $\Phi_j$ defined in Eq.~\eqref{eq:phase}, we obtain the $M$-AB envelope
	\begin{equation}
		U_j(x,t)=\frac{\Phi_j U_j^{[M]}}{\prod_{i=1}^M g_j(e_i)}.
	\end{equation}
	
	\bibliography{Manakov_references}
\end{document}